\documentclass[12pt]{amsart}

\usepackage{booktabs}
\usepackage{amsmath}
\usepackage{amssymb}
\usepackage[margin=1in, top=1.5in]{geometry}
\usepackage[slantedGreek]{mathptmx}
\usepackage{url}
\usepackage{stmaryrd}
\usepackage{amsthm}
\usepackage{pdfpages}
\usepackage{xcolor}
\usepackage{accents}
\usepackage{booktabs}
\usepackage[most]{tcolorbox}
\usepackage{float}

\title[General-purpose cooperativeness]{General-purpose cooperativeness and altruism in humans: elements of the mathematical framework for the Interdependence Hypothesis}

\author{Misha Perepelitsa}

\thanks{Email: mperepel@central.uh.edu, Address: Department of Mathematics, University of Houston, PGH 631, 4800 Calhoun Rd., Houston, TX.
USA}

\begin{document}

\begin{abstract}
We propose a decision-making model for joint intentionality by interpreting it as  group-mindedness at the microlevel. We apply this model  to give a formal justification of the first part of the Interdependence Hypothesis due to Tomasello et al. [Current Anthropology, 2012] which asserts that the emergence of joint intentionality evolved due to the  challenges of difficult  collaborative foraging practices among early humans, and that its evolution led to  robust collaboration and some form of altruism. 

In another application of the microlevel group-mindedness we consider the problem of establishing cooperation in high-risk-of-defection strategic conflicts and we show that the emergence of cooperation in such situations can be explained in the context of cultural group selection as the result of adaptive learning provided the individuals use co-utilities with the level of cooperativeness above a certain limit.

\end{abstract}


\maketitle

\section{Introduction}
Tomasello et al. (2012) proposed the Interdependence Hypothesis as an explanation of the evolutionary origin and development of  widespread cooperativeness and altruism in modern humans. This theory has two parts.  The first begins with our last common ancestor, who lived in small groups (the size of a troop of chimpanzees) subsiding on gathering plants and insects and  opportunistic group hunting, and who began transitioning to more regular and organized foraging activities such as scavenging, possibly due to ecological changes to the habitat. The latter activity required more time and resources (and thus was a more challenging collaborative activity) which set the scene for selection of more efficient collaborators.  The success of this transition critically relied on three key problems associated with such collaborative task: a. sharing of spoils; b. coordination;  c. temptation to free ride. 

The second part of the theory is concerned  with the transitioning of early humans from living in small groups  to bigger societies (tribes), which was characterized by the development of 
the cognitive ability of group-mindedness which improved  coordination and collaboration with relative strangers and led to the formation of social conventions, norms, and institutions.

In this paper, we consider the coordination step from the first part of the Interdependence Hypothesis with several goals in mind. First, we would like to build a mathematical model for the decision-making that accounts for joint attention and perspective taking. Our starting point is a utility function that an individual uses to select between alternatives. Since joint attention expresses itself as  ``group-mindedness'' between individuals and an object of cooperation (Tomasello, 2014, 2019), we conjecture that each individual, when faced with a collaborative opportunity, aggregates her own payoffs and the payoffs of her collaborator into a new utility function (called co-utility) in manner suggested by the theory of social choice in economics. The new utilities, see formula \eqref{def:cou} below, contain a parameter $\lambda,$ called  the strength of cooperation, which gives a weight of the collaborator's payoff in one's own utility.  This notion effectively captures the group-mindedness at the microlevel of two-person interaction.

Consequently, we will apply co-utilities to determine how strategies are selected in cooperative opportunities described by the stag-hunt games, since such games capture the essence of strategic conflict between cooperation and defection (Tomasello et al., 2012; Skyrms, 2007).
  
 Given a stag-hunt game, which determines fitness points for acting according to either cooperation or defection, we assume that it is internalized by the individuals who use co-utilities to evaluate outcomes. This transforms the game into another stag-hunt game for internal decision-making. We will argue, using the result of Kandori et al. (1995), that the decisions made by individuals follow a risk-dominant strategy (notion due to Harsanyi and Selten,1988).

Our second goal is the explanation of  the mechanism for the evolutionary origin of co-utility and better overall cooperativeness in a population that it brings with it.  The main assumption here is that  the strength of cooperation $\lambda,$ included as a parameter in co-utility, is under the control of natural selection.  The key observation is that the set of stag-hunt games that are played cooperatively is bigger for larger values of strength of cooperation $\lambda,$  with more risky games included in that set. At this point, it will be important to distinguish between a stag-hunt game which assigns fitness points to participants, and a corresponding decision-level stag-hunt game for intrapersonal choice, i.e. the perception of the game by individuals. 

We  propose  that the evolution to a higher level of strength of cooperation by means of  a ratcheting-up mechanism. It starts with neutral drift that moves cooperativeness to a higher level, and is followed by 
by directed evolution to cooperative ESS in newly  acquired new cooperative practices (new stag hunt games) or  in the some of the currently used practices, played non-cooperatively. The latter step  prevents a drift in the direction of decreasing levels of cooperation and leads to accumulation of the level of cooperativeness.  As we'll see in section \ref{sec:evolution}, the process is slightly more involved and includes a phase-transition-type phenomenon.
The novelty of this approach lies in considering the effect of evolving co-utilities on a set of stag-hunt games practiced by groups of early humans, rather than studying the cooperation in a context of one particular game.   

With respect to other cooperation/defection conflicts,  the process considered above  will not evolve to cooperation in a defection-dominant game such as Prisoner's Dilemma (PD), because cooperation is not an ESS.  Group selection (Maynard Smith, 1964;  Bowels, 2006; Traulsen and Nowak, 2006), but more likely,  Cultural Group Selection due to Richerson and Boyd (2005) that accounts for  different forms of reciprocation (Axelrod, 1984;  Nowak and Sigmund, 1998;  Sigmund, 2012), can explain emergence of cooperation (altruism) in such situations.

Selection of any type, cultural or gene, needs as a precondition, the selected trait to be expressed in a population in a first place. Typically, this is attributed to mutations. Another mechanism, especially relevant to humans with their developed cognitive abilities, can be all-purpose learning, which is a part of  the adaptation suit. Thus, in section \ref{sec:PD}, we consider an application of co-utilites to the situation where individuals with micro-local group mindedness, interact and adaptively  learn (Herrstein 1970;  Harley, 1981; Sutton and Barto, 1998; Hauert and Stenull 2002) in a repeated plays of Prisoner's Dilemma.  Interpreting $\lambda$ as an unconditional probability to cooperate (unconditional altruism) and  subjecting learning to random noise and memory effects, we show  that players can learn to cooperate, on average, at higher levels than their unconditional level of altruism, provided that the latter is greater than a certain threshold.  Considered in the context of cultural group evolution this shows that high levels of cooperation in a population of individuals with joint conventionality (co-utilities) can be achieved by progressive ``normalization'' of the default levels of cooperation.


\section{Evolution of microlevel group-mindedness}
\subsection{Co-utility}
Tomasello (2014, 2019) stresses the cognitive ability of taking and coordinating different perspectives as one of the main characteristics distinguishing modern humans from great apes. This cognitive ability is responsible for the development of collaboration between humans at levels far beyond those in great apes.

In the ontogenetic process of modern humans, perspective taking starts with the formation of ``joint attention'' towards an object between a child and an adult (Tomasello, 1995), which later develops into the sharing of  experiences, with abilities to align, coordinate and exchange perspectives (Tomasello 2019), thus allowing humans to   know things jointly as ``we.'' In Tomasello et al. (2012), as the part of the Interdependence Hypothesis, authors identify joint attention as the key cognitive element of the evolution of modern humans that made it possible for early humans to cooperate in risky strategic situations.

A simple model of joint attention and coordination can be built on the concept of utility (internalized scale of payoff valuation).  The joint attention can be expressed as the fact that each 
individual is aggregating her own utility with the utility of her partner, leading to ``group thinking'' at an individual level. The use of aggregated preferences is classical in Economics (and Game Theory) in the field of Group Decision-Making or Social Choice. 

Thus, suppose that we have two individuals, one with utility function $u$ and the other with utility function $v,$ that they use for  valuation of  outcomes that are not related to joint attention. When they act in a joint enterprise involving joint attention  and  the actions they select provide payoffs $a$ and $b,$ to the first and the second individual, respectively, we suppose that they use different scales and evaluate the actions in the enterprise according to the new utilities 
\begin{equation}
\label{def:cou}
u_{\lambda} = (1-\lambda)u(a) + \lambda u(b),\quad v_{\lambda} = (1-\lambda)v(b) + \lambda v(a),
\end{equation}
where parameter $\lambda$ (the strength of cooperation) ranges in the interval $[0,1].$ The definition reflects the fact that ``our perspectives of the joint activity become somewhat aligned if each of us, at least partially, attends to the interests of the other.'' We will refer to \eqref{def:cou} as co-utilities (cooperative utilities). In the language of Economics, when acting in this way, the individuals are exercising group-thinking at the microlevel. When a cooperative opportunity (symbolized by a 2-person game) is presented to the individuals using  joint attention, they will evaluate 
action alternatives (strategies) using \eqref{def:cou} and select actions according to a suitable  solution of strategic conflict (a solution in the 2-person game).  Diagram \eqref{pathway1} below illustrates the sequence of steps in decision making at the  stag-hunt game from Tables \ref{game:stag} and \ref{game:stag_u} that we will use below.

With this approach we can analyze the effect of the strength of cooperation on the behavior selection in {\it the totality of cooperative activities} (expressed as stag-hunt games), rather than studying cooperation as a  strategy in one particular game. Consequently, our argument for the evolutionary origin of co-utilities will be grounded on the fact that  larger values of $\lambda$ in \eqref{def:cou} result in individuals selecting cooperation for {\it a larger set} of stag-hunt games, progressively including more risky games, thus in its effect increasing the cooperativeness of the individuals in the population.

Let us remark that by introducing co-utilities e are not reducing (or making equivalent) joint attention to them. Rather, we are suggesting that they are a part of the psychological state of the joint attention, which might even had precluded the development of the latter.

\begin{table}[h]

\begin{tabular}{@{}lll@{}}
\toprule
 & C \hspace{20pt} & D\\
\midrule
 C \hspace{10pt} & (s,\,s)                         & (d,\,r) \\
D \hspace{10pt} & (r,\,d)      & (h,\,h) \\
\bottomrule
\end{tabular}
\vskip 20pt
\caption{Stag Hunt game (SH). Pairs of payoffs, for  player I (left) and player II (top), respectively, with
$s>r>h>d.$ 
When both players cooperate, both have the same payoff $s,$ highlighting an implicit assumption of the fairness of sharing of spoils.}
\label{game:stag}
\end{table} 

\begin{table}[h]
\centering
\begin{tabular}{@{}lll@{}}
\toprule
 & C \hspace{20pt} & D\\
\midrule
 C \hspace{10pt} & u(s)                         &  $(1-\lambda)$u(d) + $\lambda$u(r)  \\
 D \hspace{10pt} & $(1-\lambda)$u(r) + $\lambda$u(d)      & u(h) \\
\bottomrule
\end{tabular}
\vskip 20pt
\caption{Stag Hunt game $\rm SH(\lambda).$ Table shows cooperative decision-making utilities of player I in SH-game from Table \ref{game:stag}. Player I uses utility function $u(\cdot)$ for internal representation of the game. Cooperativeness level of player I is measured by $\lambda\in[0,1].$ The ultimate cooperator, $\lambda=1,$ has her opponent's (player II) best interests in mind, measured by her own utility.}
\label{game:stag_u}
\end{table}

\subsection{Stag Hunt game} Following Skyrms (2007) and Tomasello et al. (2012) we represent a cooperative opportunity as a stag-hunt (SH) game, Table \ref{game:stag}, where the payoffs are interpreted as fitness points.  SH-game has two symmetric ESSs, $C$ and $D$ and an unsteady mixed Nash equilibrium $x_0C+(1-x_0)D,$
where 
\[
x_0 = \frac{h-d}{s-r+h-d}\in(0,1).
\] 
We consider this game as a complete information game where all participants are aware of the rules and outcomes as they participate in that activity on a regular basis and have the necessary cognitive abilities. 

The corresponding decision-making game, when individuals use $u_\lambda$ utilities, is given in Table \ref{game:stag_u}. We will assume that $u_\lambda$ is an expected utility (linear on mixed strategies): for example if the first player uses $C$ and the second uses mixed strategy $\alpha C + (1-\alpha)D,$ then the first player co-utility 
\[
u_\lambda\left(C(\alpha C + (1-\alpha)D)\right)=\alpha u_\lambda(CC) + (1-\alpha)u_\lambda(CD),
\]
 and, for simplicity, we select the original (non-cooperative) utility function for both players to be $u(s)=s,$ i.e., the identity function.   For a SH-game, and $\lambda\in[0,1],$ the new game written in cooperative utilities $u_\lambda$ is just another SH-game,  when $\lambda < \frac{h-d}{r-d}$, or a game with the dominant strategy $C.$ In either case, we will call it $\rm SH({\lambda})$ game.

\begin{figure}[htbp]
\centering
\includegraphics[trim=10 510 80 10, scale=0.8]{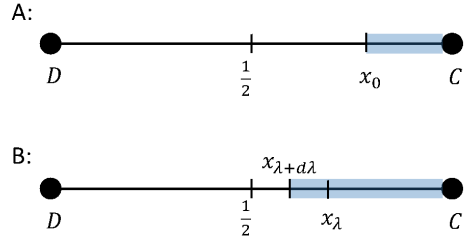}
\caption{Risk dominance in stag-hunt games. A: shows the basin of attraction for cooperation $C$
in SH-game from Table \ref{game:stag}. B: shows the basin of of attraction in $C$ in game $\rm SH(\lambda+d\lambda),$ widh $d\lambda>0,$ see Table \ref{game:stag_u}.  The basin of attaction of $C$ increases with $\lambda.$ In all cases shown in this figure, $D$ is the risk dominant and is selected as a preferred behavior.
\label{fig:basin}}
\end{figure}

The interaction cycle between two participants  proceeds in the following sequence
\begin{equation}
\label{pathway1}
\fbox{\begin{tabular}{c}
Cooperative opportunity: \\
SH game\end{tabular}}\Longrightarrow \fbox{\begin{tabular}{c}Internal representation:\\
 $\rm SH(\lambda)$ game\end{tabular}}
\underset{\rm action}{\Longrightarrow}\fbox{\begin{tabular}{c}Fitness points:\\ SH game\end{tabular}}
\end{equation}
\vskip 10pt
which will be completely specified once the rules of selection of the actions in the last step are prescribed. For that, if the new $\rm SH(\lambda)$ game has a dominant strategy (can only be $C$), it will be selected, resulting in cooperative choice $C.$  If  $\rm SH(\lambda)$ game has two ESSs, we assume that the decision is made  based on the least risky Nash equilibrium, i.e., where the participants use  the principle of risk dominance (Selten and Harsanyi, 1998), which says that the proportion of the preferred strategy in the Nash mixed equilibrium is less than 1/2: facing an opponent that plays $C$ or $D$ with probability 1/2, the best response is to play this preferred strategy. In the case of $\rm SH(\lambda)$ game,
\begin{equation}
\label{choice_rule}
\rm
\mbox{$C$ is selected over $D$  when $\dfrac{h-d-\lambda (r-d)}{s-r+h-d}<\dfrac{1}{2},$}
\end{equation}
and vise versa. We label by $x_\lambda$ the fraction on the left-hand side in \eqref{choice_rule}, see Figure \ref{fig:basin}. The measure of the risk  expressed in this formula by quantities $h-d>0$ and $s-r>0.$ The first is the incentive to defect when your opponent cooperates, and the other is the incentive to to cooperate if your opponent cooperates. Another way to interpret rule \eqref{choice_rule} is by noticing that it implies that the basin of attraction of the preferred state in evolutionary dynamics is larger than the basin of attraction of the other state. More importantly, it was shown by Kandori et al. (1995) that the risk dominant strategy is the only limiting state of any ``Darwinian'' process stable with respect to random ``mutations'' (noise in strategy choice) even if mutations are biased toward the opposite state. In that paper, by ``Darwinian'' authors mean a process in which population frequency of a better performing strategy increases.  This result gives a strong indication that the risk dominance choice rule \eqref{choice_rule} (or rather cognitive structures responsible for such choice)  could have been selected in evolution. We will assume this hypothesis as one of the preconditions for the evolution of co-utilities.  A special role in the theory described just below will be played by the borderline case $x_\lambda=\frac{1}{2}.$  We will argue that the decisions made at this point depend on the state of the system at earlier stages of the evolutionary process, followed by an irreversible ``phase transition'' from defection to cooperation.

\subsection{Evolutionary process}
\label{sec:evolution}
Suppose that the group is regularly  engaged in a number of cooperative opportunities (modeled as stag-hunt games). In some of them the behavior is to cooperate, $C.$ We denote them by 
\[
\mathcal{A}_c = \left\{\mbox{Set of SH games played cooperatively} \right\},
\]
while in other the behavior is $D,$  
\[
\mathcal{A}_d = \left\{\mbox{Set of SH games played non-cooperatively} \right\}.
\]
We are going to describe the evolution of $\lambda$ from zero to some postivie level. In this process $\lambda$ increases in small (infinitesimal) increments trough a ratcheting mechanism that prevents it is going backward. 

Let $\lambda$ be the current level of cooperation in the group, according to \eqref{def:cou}.
According to the selection rule \eqref{choice_rule} is a game is in $\mathcal{A}_c$ then its $x_\lambda$--value is less than $\frac{1}{2},$ and if the game is in $\mathcal{A}_d$ then $x_\lambda>\frac{1}{2}.$
Suppose a mutation occurred and some members of the group now have slightly higher values of cooperation: $\lambda+d\lambda.$ 

First, we notice that the mutation can spread by neutral drift.
Indeed, if  $x_\lambda$--value of an SH-game from $\mathcal{A}_c$ or $\mathcal{A}_d$ is more than $d\lambda$ units away from $\frac{1}{2},$ then the decision to cooperate or not in this game (based on rule \eqref{choice_rule}) does not change with new co-utility, see Figure \ref{fig:basin}. More than that, if the game is from $\mathcal{A}_c,$ it will be played cooperatively for any positive amount of change $d\lambda.$ 

But it is also true that a situation may arise when $x_\lambda$--value of  a game from $\mathcal{A}_d$ is shifted in the view of mutants to the borderline case: $x_{\lambda+d\lambda}=\frac{1}{2}.$ For those mutants both $C$ and $D$ are now equally good, the decision will be based on some other principles, which we assume is related to ``conformism'' or ``traditionalism,'' with mutants continuing to behave the same way as the majority or as they were taught. In this way the number of mutants may increase by neutral drift in this case too.

Next step is the ratcheting-up mechanism that  can  manifest itself twofold.  Suppose that  the group acquires new cooperative practices. A new cooperative opportunity (new stag-hunt game) might be discovered by the  group that now will be played cooperatively by mutants with $\lambda+d\lambda$, but not by members of the  group  the cooperation strength $\lambda.$ A sufficiently large number of mutants will eventually out-evolve the remaining orthodox, since cooperation is an ESS in a stag-hunt game. The latter property will also insure the stability of mutants, preventing the neutral drift to decrease the level of the cooperation strength.  This also explains why using stag-hunt as opposed to Prisoner's Dilemma is essential to the argument. 
 
Consider now a stag-hunt game from $\mathcal{A}_d$ for which the new level of the cooperation strength is the borderline case $x_{\lambda+d\lambda}=\frac{1}{2}.$ Recall that we argued that the group keeps using $D$ in this case even if mutants compose the majority or all of the group. Note that while $C$ and $D$ are equally risky, $C$ produces higher fitness to participants since in any stag-hunt game, $CC$ dominates $DD.$ It seems likely that through experimentation or noise the better cooperative strategy will be discovered and adopted by majority. We will call this phenomenon the phase transition at $\frac{1}{2}$--value.  Notice that the process is irreversible, proceeding from non-cooperation to cooperation, also preventing backward mutations from spreading.  With the completion of the phase transition, the proportions of people adopting $C$ will be somewhere in the basin of attraction of $C$ in the original $\rm SH$ game (the one that determines the fitness points for strategies), which will lead to the complete takeover of that group by mutants.  

In this way, by accumulating neutral drifts in a preferred direction and by accumulating new cooperative practices (updating $\mathcal{A}_c$ and $\mathcal{A}_d$), the group will  move to distinctly higher (not just infinitesimally) values of cooperation strength, thus adopting a cooperation in more risky situations.

We restricted the above analyses to the context of a stag-hunt game. Mentioned above Prisoner's Dilemma is another game routinely used as a paradigm of cooperation-defection conflict, often in the context of altruism.
Since co-utilities in principle can be used in any type of joint activities, we can stipulate its application for PD games. Here, however, the defection dominance in PD makes defection the only ESS to which the group will always gravitate towards, even if  a mutation produces high values $\lambda$ in \eqref{def:cou} and cooperation is selected as a better alternative by co-utility, Table \ref{game:PD2}.

\begin{table}[h]
\centering
\begin{tabular}{@{}lll@{}}
\toprule
 & C \hspace{20pt} & D\\
\midrule
 C \hspace{10pt} & $b_0+b-c$                         & $b_0-c$ \\
 D \hspace{10pt} & $b_0+b$     & $b_0$ \\
\bottomrule
\end{tabular}
\vskip 20pt
\caption{Prisoner's Dilemma with benefit $b>0$, cost of altruistic behavior $0<c<b$ and baseline of $b_0\geq c$ fitness points.}
\label{game:PD1}
\end{table} 

\begin{table}[h]
\centering
\begin{tabular}{@{}lll@{}}
\toprule
 & C \hspace{20pt} & D\\
\midrule
 C \hspace{10pt} & $b_0+b-c$                         & $b_0+\lambda b -(1-\lambda)c$ \\
 D \hspace{10pt} & $b_0+(1-\lambda)b-\lambda c$      & $b_0$ \\
\bottomrule
\end{tabular}
\vskip 20pt
\caption{Prisoner's Dilemma. Decision-making utilities with cooperation level $\lambda.$ With $\lambda>c/(b+c),$ $C$ dominates $D,$ but cooperation can not evolve because $D$ is the only stable ESS in the original PD game in Table \ref{game:PD1}}
\label{game:PD2}
\end{table} 


\section{Learning to cooperate in defection-dominated games}
\label{sec:PD}
As we noted above, the evolution of the internal co-utility towards more altruistic levels is not sufficient for establishing full or partial cooperation in defection-dominated games such as Prisoner's Dilemma. 

In this section we would like to show that individuals engaged in a repeated plays of PD game can learn 
to cooperate even if they are mildly altruistic. More precisely, the can learn to cooperate to significantly higher levels than suggested by their internal altruistic levels. As the learning can take a modest number of interactions, the individuals can learn to cooperate in a span of one generation, and after a few generations the practice can spread across the group and adopted as a new norm.

To build the model we will interpret co-utility $(1-\lambda)u(a) + \lambda u(b)$ as an expected value of a random choice, where an individual using the ``selfish'' utility of her own payoff $u(a)$ with probability $1-\lambda$ and the altruistic utility of the opponent's payoff $u(b),$ with probability $\lambda.$ Thus, we let the players behave altruistically (cooperatively) with probability $\lambda$ and competitively with probability $1-\lambda.$ When in cooperative mode, a player always chooses strategy $C$ because it maximizes the payoff of the opponent in PD game. 

The learning algorithm will be implemented through a general-purpose reinforcement learning, when the propensities to cooperate and defect are updated with the current payoffs to $C$ or $D$ actions, according to the game in Table \ref{game:PD1}. The probability to play a strategy at the next round is a fraction of the propensity for that action to the sum of propensities for both actions. 

We assume that learning is subject to a memory effect, expressed by a parameter $\mu>0,$ which we select such that a current value of the propensity to $C$ or $D$ is reduced by order of $10^{-1}$ after about 10 interactions (each individual effectively remembers around last 10 interactions). 

The interaction will be subjected to noise with rate $\eta,$ so that from time to time an intended strategy 
is implemented as the alternative one. In $N$ interaction there is on average $\eta N$ mistakes in the implementation.  Moreover, we assume that both individuals are  initially are skeptical about the cooperation, with zero initial chances to cooperate. 

\begin{figure}[H]
\centering
\includegraphics[clip, scale=1]{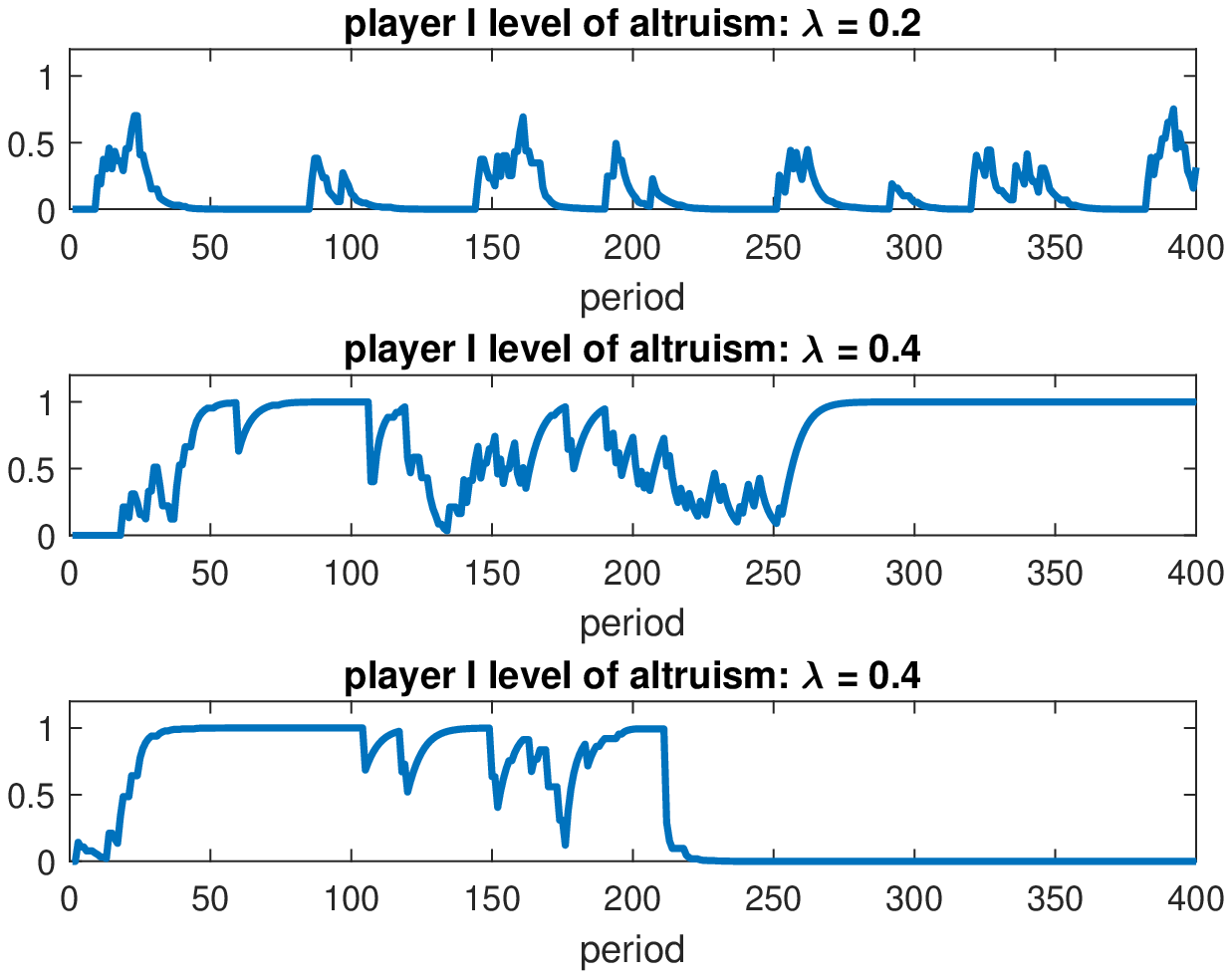}
\caption{
\label{fig:learn1}Learning dynamics in Prisoner's Dilemma. Top: simulated learned probability to cooperate of one of the players when the default level of cooperation (altruism) of both players $\lambda=0.2.$ Plot shows first 400 periods of interaction. Middle: same as top but with the default level of cooperation (altruism) of both players $\lambda=0.4.$ The process repeatedly reaches full cooperation and locks in there until the noise pushes it away.  Bottom: simulated learned probability to cooperate of player I, when player II switches from learning to defection after period 200.}
\end{figure}

The dynamics of this learning process is illustrated in Figure \ref{fig:learn1}, for several values of $\lambda.$ 
The mutual cooperation pushes players toward full cooperation.  With higher values $\lambda,$ the players lock in (lock-in is due to short memory), until noise pushes them out. At which point the cooperation completely disintegrates. The players however cannot lock in defections because of their non-zero altruism levels, and the process repeats itself in a stationary stochastic manner, with players repeatedly keep learning and locking in full cooperation.  In this simulations we used a PD game with $b_0=c,$ so that cooperation cannot be self-reinforcing when the opponent defects, see the last panel in Figure \ref{fig:learn1}.

\begin{figure}[H]
\centering
\includegraphics[clip, scale=1]{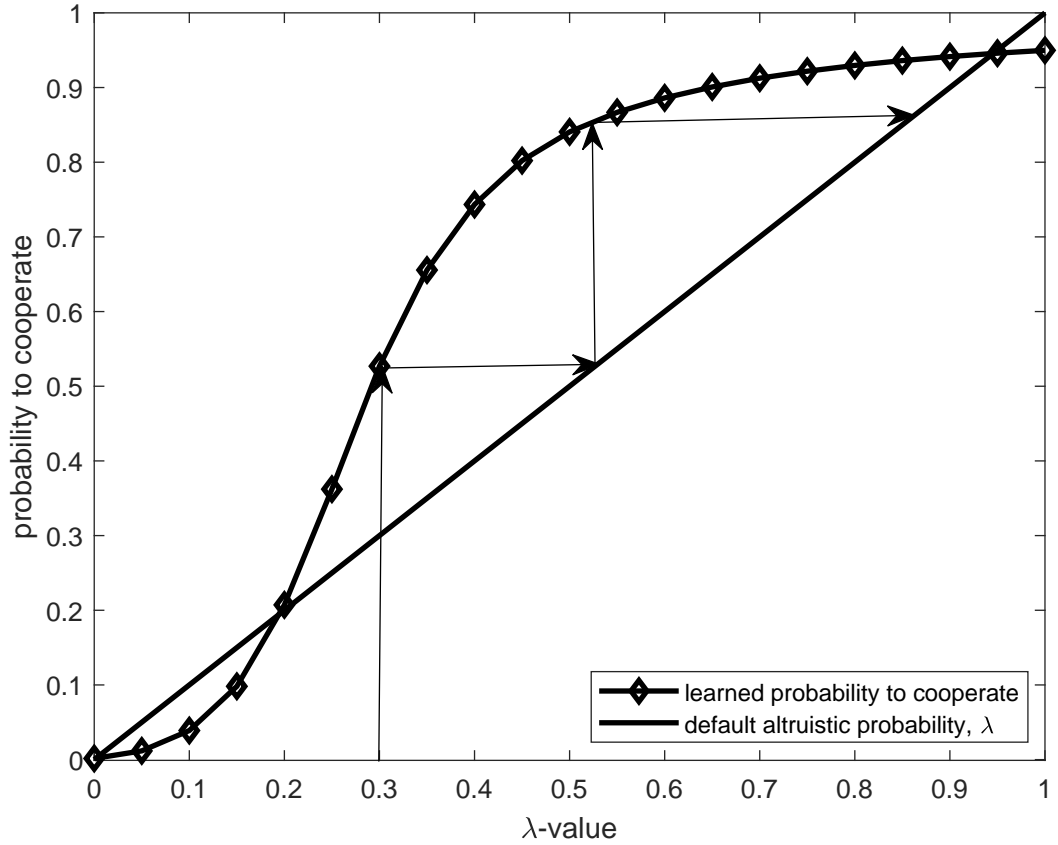}
\caption{Mean learned probability to cooperate in competitive mode as a function of default cooperation (altruism) level. The plot shows simulated learned probability to cooperate at values of $\lambda$ between 0 and 1, with increments of $0.05,$ in comparison to the default cooperation level. The value of the learned probability to cooperate is averaged over first 400 periods and $\rm10^5$ different paths (simulations). In the range $\lambda\in (0.2,\,0.95),$ learning overshoots the default level of altruism.  Arrows indicate two steps of   culturally evolved levels of default probability to cooperate (altruism) starting at value $\lambda_0=0.3.$ Small values of $\lambda$ ($<0.2$) lead to the process of diminishing cooperation.
\label{fig:learn2}}
\end{figure}

Figure \ref{fig:learn2} shows how much participants learn to cooperate (on average) as a function of the altruism level $\lambda,$ when they take actions in non-cooperative (competitive) mode. Note the significant range of $\lambda$ for which the learned probability is greater than the default level of altruism.

If we view this type of learning  in the context of cultural evolution, i.e., with useful practices becoming normative rules transmitted by teaching, then the learning can take the form of the processes  
\[
\fbox{\begin{tabular}{c}
Normative (default)\\ cooperation
$\lambda_n$\end{tabular}}\Longrightarrow \fbox{\begin{tabular}{c}Learning process:\\
 stationary cooperative level $\lambda_s$ \end{tabular}}
\Longrightarrow\fbox{\begin{tabular}{c}New normative \\ cooperation $\lambda_{n+1}=\lambda_s$\end{tabular}}
\]
that will eventually take cooperation to a high level. Figure \ref{fig:learn2} shows first two steps of this process  that eventually ends up the rightmost point of intersection of two graphs.   The point of the Figure \ref{fig:learn2} is to show that the learned level of cooperation can be higher than the default level, leading to a gain in cooperativeness. Notice that this property holds when $\lambda$ is greater than a certain threshold value $\lambda=0.2,$ below which the learning takes participants to defection by progressively decreasing the cooperativeness (not shown in Figure \ref{fig:learn2}).  

Thus the model illustrates that altruism can be established as a cultural norm in specific types of interactions in a group of humans with mild levels of personal all-purpose altruism in their internal co-utilities, but for this to work, participants must have co-utilities with $\lambda$ above certain limit. As we showed in the previous sections, the latter property could evolved in the context of stag-hunt games.


\section{Connection to altruism}
\label{sec:altruism}
As we noted above, co-utilities \eqref{def:cou} characterize the joint attention to a collaborative activity expressed symbolically as a stag hunt game. Presumably, at the time of origination of joint attention,
 the use of co-utilities were restricted to the situations that required  immediate attention of the interacting parties when the actual presence of the collaborator on the scene as well as the target of collaboration were needed for the activation of joint attention.  That is,  co-utilities were only used in situations when ``I will use $u_\lambda$ if and only if you use $v_\lambda.$''  This is a sort of tit-for-tat approach for the way the strategies are evaluated. With ever increasing number of co-operational links between individuals in human growing societies, co-utility might have become a useful behavioral heuristic for cooperation when strengthened by the parallel cultural development of reputation markers by the processes of cultural transmission, see  Richerson and Boyd (2005),  Henrich and Henrich (2007). Moreover, acting according to co-utility $u_\lambda$ without any reference to what your potential cooperator might do acquires high normative (moral) value due to its extreme usefulness. It also seem plausible that the usage of co-utility had diffused from the realm of stag-hunt games, where it originated, to other types of strategic conflict.  In such contexts we can speak of $u_\lambda$ as an all-purpose altruistic utility, with $\lambda$ expressing the degree of altruism. The extreme value of $\lambda=1$ certainly refers to what one would call altruistic behavior.




Here we align with the thesis appeared in Tomasello et al. (2012)  that uniquely human way of cooperation  was not caused by any form of task-specific altruism that proto-humans acquired earlier, but developed independently (co-evolved with joint intentionality) and carried with it its own form of altruism:  altruism as an investment in a potential collaborator (Tomasello et al., 2012), or altruism as an all-purpose behavioral heuristic.

This concept naturally complements the existing theories of  task-specific altruistic behaviors in animals and humans by  kin selection, due to Haldane (1955) and Hamilton (1964), or  group selection, Maynard Smith (1964),  Bowels (2006) and Traulsen and Nowak (2006).


\section{Conclusions}
In this paper we considered a mathematical model of decision-making under joint attention 
and used it to describe transitioning from low-level (and low risk)  or opportunistic  cooperation in great apes  to more systematic and robust cooperation in humans through the evolution of joint attention, expressed, using co-utilities,  as microlevel group-mindness.

Group thinking has a direct effect on the cooperativeness of the individuals, as ``cooperation'' strategy gets selected in more cooperative opportunities. We term such cooperation (and altruism derived from it) the all-purpose cooperativeness, since it is not tied to one particular situation of strategic conflict, but rather works at the level of evaluating outcomes for decision making. 

The model for the evolutionary origin and change in co-utilities involves three  distinct evolutionary ``dimensions'' or steps. The first, is the  evolution of cognitive abilities to select risk dominant strategy in a stag-hunt game, and to select strategy with higher fitness among all strategies with equal risk. They serve as  preconditions for the second dimension -- the neutral drift in the strength of cooperation $\lambda.$ The third step is the directed evolutionary convergence to the cooperative ESS in newly acquired cooperative activities (or re-considered old, non-cooperative activities). The last step is the ratcheting mechanism that accumulates the positive increases in the level of cooperativeness and prevents it from moving backwards. 

The convergence in the last step could be furthermore strengthened by cooperative mutants selecting against non-cooperative individuals by not choosing them as collaborative partners, see Tomasello et al. (2012). In addition, the neutral drift could be not neutral, but biased towards mutants, as we mentioned earlier. However, even in the conservative scenario considered in the paper, the model predicts the evolution of higher cooperative co-utilities.

As the evolution of co-utilities depends on discovering new opportunities of cooperation, it most likely co-evolved with other group practices, such as food sharing, cooperative breeding, and teaching, and  cognitive abilities for enhanced  communication and coordination skills (Henrich and Henrich 2007; Laland 2017). 

The question can be asked why didn't $\lambda$ evolved all the way to unconditional altruism level of $\lambda=1.$ In part, it might be due to factors mentioned in the previous paragraph, which also promote cooperation. Additionally, the adaptive learning that we used in section \ref{sec:PD} to explain the onset of cooperation in Prisoner's Dilemma-type interactions, in the process of cultural evolution, can be applied in the context of stag-hunt games, where it certainly will be even more efficient.  Thus, while the evolution from $\lambda=0$ to a positive $\lambda$ seems to be essential, the later development could slow down and completely stop way before $\lambda$ reaches the unconditionally altruistic level, because it was learned and standardized as a norm.


\section{Appendix}

\subsection{Reinforcement learning model}
We assume that two players engaged in repeated play of Prisoner's Dilemma from section \ref{sec:PD}. Let $\lambda$ be their default level of altruism, $\mu$ be the memory factor and $\eta$ -- noise rate. 
Let $P_{i,c}^n$ and $P_{i,d}^n$ be the payoff to player $i=1,2$ at period $n$ if she used strategy $C,$ or $D,$ respectively (when strategy is not used, the payoff is set to zero).
The propensity to cooperate and defend for each player are updated according to the reinforcement rule (Harley, 1981; Sutton and Barto, 1998; and a related model of Hauert and Stenull 2002):
\[
S_{i,c}^{n+1} = \mu S_{i,c}^n + P_{i,c}^n,\quad S_{i,d}^{n+1} = \mu S_{i,d}^n + P_{i,d}^n.
\]
At period $n,$ with probability $\lambda$ each player selects the altruistic outcome $C.$ Otherwise, player $i$ selects $C$ with probability
\[
\frac{S_{i,c}^n}{S_{i,c}^n+S_{i,d}^n}, \quad i=1,2,
\] 
and $D$ with the complimentary probability.  After the strategy is selected it is subject to random noise with probability $\eta$ of switching to another strategy. In the simulations the following values were used: 
$\eta= 0.03,$ $\mu= 0.8,$ initial propensities $S_{i,c}^0(0) = 0; S_{i,d}^0(0)=10,$ in a PD game with values  $b=1.2$, $c=b_0=1$.


\newpage


\end{document}